\documentclass[12pt,a4paper]{article}
\usepackage{amsmath}
\usepackage{latexsym}
\makeatletter
\def\rddots{\mathinner{\mkern1mu\raise\p@%
    \vbox{\kern7\p@\hbox{.}}\mkern2mu%
    \raise4\p@\hbox{.}\mkern2mu\raise7\p@\hbox{.}\mkern1mu}}
\makeatother
\begin{document}

\title{\sl A Multidimensional Analogue of the \\
Simpson's Formula of Integral}
\author{
  Kazuyuki FUJII
  \thanks{E-mail address : fujii@yokohama-cu.ac.jp }\\
  Department of Mathematical Sciences\\
  Yokohama City University\\
  Yokohama, 236--0027\\
  Japan
  }
\date{}
\maketitle
\begin{abstract}
  The Simpson's formula is obtained by approximating 
  the integral of a function on some interval by the integral 
  of the quadratic polynomial determined by the function. 
  However, a multidimensional analogue of the formula has not  
  been given as far as we know. 
  
  In this paper such a formula is given. Our formula is simple 
  and beautiful, so it may be convenient in Mathematics or 
  Mathematical Physics.
\end{abstract}
%

%
%
%
%
\section{Introduction}
Let $y=f(x)$ be a continuous function defined on the interval 
$[a,\ b]$.  We want to calculate the integral $\int_{a}^{b}f(x)dx$.  
However, it is not easy to calculate the integral explicitly, so 
to make some approximation is a realistic way. We recall 
the Simpson's formula \footnote{The author uses the 
terminology {\bf Simpson's formula} in place of Simpson's rule 
because it is nothing but an approximate formula (or method)} 
, see for example \cite{WM}.

We divide  $[a,\ b]$ into equal $2n$ subintervals and set 
$\Delta=\frac{b-a}{2n}$, and $a=x_{0}$, $b=x_{2n}$.

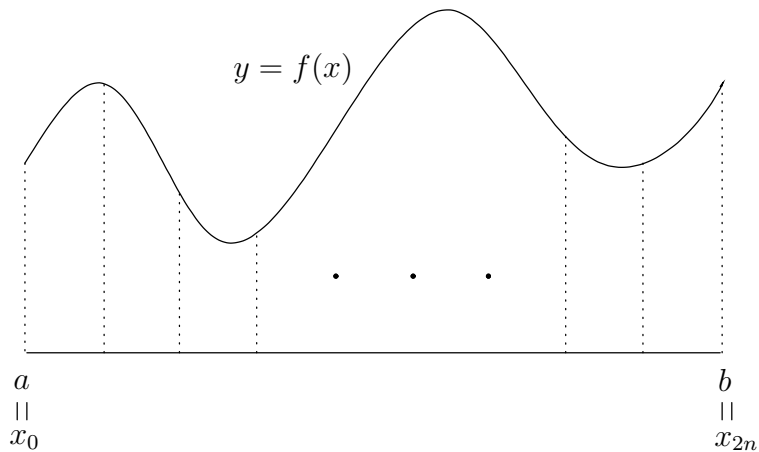
\vspace{3mm}
\begin{center}
 \input{integral.fig}
\end{center}

\vspace{2mm}
Now we study the integral on small interval $[x_{2k-2},\ x_{2k}]$ 
in the following. 

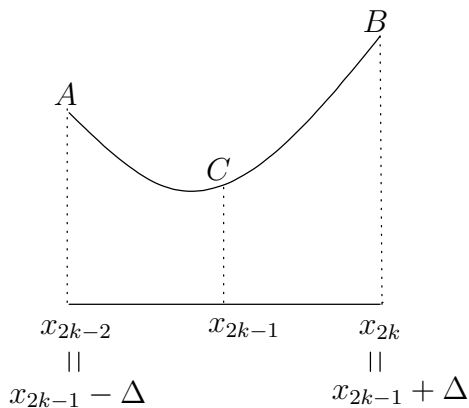
\vspace{2mm}
\begin{center}
 \input{simpson.fig}
\end{center}

\par \vspace{3mm} \noindent
For
\[
y_{2k-2}=f(x_{2k-2}),\quad y_{2k-1}=f(x_{2k-1}),\quad y_{2k}=f(x_{2k})
\]
we set
\begin{eqnarray*}
A&=&(x_{2k-2},\ y_{2k-2})=(x_{2k-1}-\Delta,\ y_{2k-2}), \\
C&=&(x_{2k-1},\ y_{2k-1}), \\
B&=&(x_{2k},\ y_{2k})=(x_{2k-1}+\Delta,\ y_{2k})
\end{eqnarray*}
for simplicity.

A quadratic polynomial
\[
y=px^{2}+qx+r
\]
passing through three points $A$,\ $B$,\ $C$ is determined 
uniquely. Namely, we have only to solve the simultaneous equations
\[
\left\{
\begin{array}{l}
\left(x_{2k-1}-\Delta\right)^{2}p+\left(x_{2k-1}-\Delta\right)q+r=y_{2k-2}, \\
x_{2k-1}^{2}p+x_{2k-1}q+r=y_{2k-1}, \\
\left(x_{2k-1}+\Delta\right)^{2}p+\left(x_{2k-1}+\Delta\right)q+r=y_{2k}
\end{array}
\right.
\]
and the result is
\begin{eqnarray*}
p&=&\frac{y_{2k}-2y_{2k-1}+y_{2k-2}}{2\Delta^{2}}, \\
q&=&\frac{y_{2k}-y_{2k-2}}{2\Delta}-
2\frac{y_{2k}-2y_{2k-1}+y_{2k-2}}{2\Delta^{2}}x_{2k-1}, \\
r&=&y_{2k-1}-\frac{y_{2k}-y_{2k-2}}{2\Delta}x_{2k-1}+
\frac{y_{2k}-2y_{2k-1}+y_{2k-2}}{2\Delta^{2}}x_{2k-1}^{2}.
\end{eqnarray*}
Therefore the function $y=f(x)$ on the interval $[x_{2k-2},\ x_{2k}]$ 
is approximated by $y=px^{2}+qx+r$ like

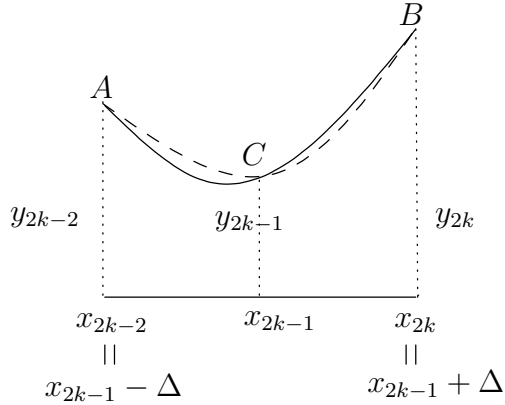
\vspace{-3mm}
\begin{center}
 \input{simpson-approx.fig}
\end{center}

\vspace{3mm}
Then an important formula is obtained
\[
\int_{x_{2k-1}-\Delta}^{x_{2k-1}+\Delta}\left(px^{2}+qx+r\right)dx
=\frac{\Delta}{3}\left(y_{2k-2}+4y_{2k-1}+y_{2k}\right)
\]
because
\begin{eqnarray*}
\mbox{LHS}
&=&
\frac{p}{3}\left\{(x_{2k-1}+\Delta)^{3}-(x_{2k-1}-\Delta)^{3}\right\}
+
\frac{q}{2}\left\{(x_{2k-1}+\Delta)^{2}-(x_{2k-1}-\Delta)^{2}\right\}
+
2r\Delta \\
&=&
\frac{2}{3}p\Delta^{3}+2(px_{2k-1}^{2}+qx_{2k-1}+r)\Delta \\
&=&
\frac{2}{3}p\Delta^{3}+2y_{2k-1}\Delta \\
&=&
\frac{\Delta}{3}\left\{2\Delta^{2}\times \frac{y_{2k}-2y_{2k-1}+
y_{2k-2}}{2\Delta^{2}}+6y_{2k-1}\right\} \\
&=&
\frac{\Delta}{3}(y_{2k-2}+4y_{2k-1}+y_{2k})=\mbox{RHS}.
\end{eqnarray*}

From this we have the famous Simpson's formula
\begin{eqnarray*}
\int_{a}^{b}f(x)dx 
&=&
\sum_{k=1}^{n}\int_{x_{2k-2}}^{x_{2k}}f(x)dx
=
\sum_{k=1}^{n}
\int_{x_{2k-1}-\Delta}^{x_{2k-1}+\Delta}f(x)dx \\
&\approx& 
\sum_{k=1}^{n}
\int_{x_{2k-1}-\Delta}^{x_{2k-1}+\Delta}\left(px^{2}+qx+r\right)dx \\
&=&
\sum_{k=1}^{n}
\frac{\Delta}{3}(y_{2k-2}+4y_{2k-1}+y_{2k}) \\
&=&\frac{b-a}{6n}\left\{y_{0}+4(y_{1}+y_{3}+\cdots +y_{2n-1})
+2(y_{2}+y_{4}+\cdots +y_{2n-2})+y_{2n}\right\}
\end{eqnarray*}
because $\Delta=\frac{b-a}{2n}$.

\section{Generalized Simpson's Formula}
Let $D$ be an $n$--dimensional cuboid
\[
D=[a_{1}, b_{1}]\times [a_{2}, b_{2}]\times \cdots [a_{n}, b_{n}] \subset {\bf R}^{n}
\]
and $f$ be a continuous function on $D$. What we do is to calculate 
the multidimensional integral
\begin{equation}
\int\int \cdots \int_{D}f(x_{1}, x_{2}, \cdots , x_{n})dx_{1}dx_{2} \cdots dx_{n}.
\end{equation}
However, it is almost difficult, so we must satisfy only by approximating it 
as shown in the introduction. 
For that purpose we rewrite $D$ as follows.
\begin{equation}
D=
[\alpha_{1}-\Delta_{1}, \alpha_{1}+\Delta_{1}]\times 
[\alpha_{2}-\Delta_{2}, \alpha_{2}+\Delta_{2}]\times 
\cdots \times 
[\alpha_{n}-\Delta_{n}, \alpha_{n}+\Delta_{n}]
\end{equation}
where $\alpha_{j}=(b_{j}+a_{j})/2$ and $\Delta_{j}=(b_{j}-a_{j})/2$. 
See the following figure ($n=2$).

\vspace{10mm}
\hspace{-20mm}
 \input{rectangle.fig}

\vspace{10mm}
Next we approximate $f$ by a quadratic polynomial
\begin{equation}
\label{eq:quadratic polynomial}
F(x_{1}, x_{2}, \cdots , x_{n})
=\sum_{i_{1},i_{2},\cdots,i_{n}=0}^{2}a_{i_{1}i_{2}\cdots i_{n}}
  x_{1}^{i_{1}}x_{2}^{i_{2}}\cdots x_{n}^{i_{n}},
\end{equation}
which is a natural extension in case of $n=1$.

For $j_{1},j_{2},\cdots,j_{n}\ \in\ \{-1,0,1\}$ we set
\begin{eqnarray}
\label{eq:lattice-points}
w_{j_{1}j_{2}{\cdots}j_{n}}
&=&
f(\alpha_{1}+j_{1}\Delta_{1},\alpha_{2}+j_{2}\Delta_{2},\cdots, \alpha_{n}+j_{n}\Delta_{n})
\nonumber \\
&=&
F(\alpha_{1}+j_{1}\Delta_{1},\alpha_{2}+j_{2}\Delta_{2},\cdots, \alpha_{n}+j_{n}\Delta_{n}).
\end{eqnarray}
It is of course 
\[
\sharp \{ w_{j_{1}j_{2}{\cdots}j_{n}} \}=3^{n}=\sharp \{ a_{i_{1}i_{2}{\cdots}i_{n}} \},
\]
see the figure above. Then we have

\vspace{5mm} \noindent
{\bf Formula (Conjecture)}
\begin{eqnarray}
\label{eq:fundamental-formula}
&&\int\int \cdots \int_{D}F(x_{1}, x_{2}, \cdots , x_{n})dx_{1}dx_{2} \cdots dx_{n}
\nonumber \\
&& \nonumber \\
&&=\frac{\Delta_{1}\Delta_{2}\cdots \Delta_{n}}{3^{n}}
      \sum_{ j_{1},j_{2},\cdots ,j_{n} \in \{-1,0,1\} }
      4^{\sharp\{j_{1},j_{2},\cdots ,j_{n}\}_{0}}w_{j_{1}j_{2}{\cdots}j_{n}}
\end{eqnarray}
where $\sharp\{j_{1},j_{2},\cdots ,j_{n}\}_{0}$ is the number of $0$ in 
$\{j_{1},j_{2},\cdots ,j_{n}\}$.

\vspace{5mm}
For the case of $n=2$ and $n=3$ the formula is proved in the following. 
However, the proof of the general case is left to readers.

From the formula we obtain a good approximation
\begin{eqnarray}
&&\int\int \cdots \int_{D}f(x_{1}, x_{2}, \cdots , x_{n})dx_{1}dx_{2} \cdots dx_{n}
\nonumber \\
&& \nonumber \\
&&\approx 
\frac{\Delta_{1}\Delta_{2}\cdots \Delta_{n}}{3^{n}}
  \sum_{ j_{1},j_{2},\cdots ,j_{n} \in \{-1,0,1\} }
  4^{\sharp\{j_{1},j_{2},\cdots ,j_{n}\}_{0}}w_{j_{1}j_{2}{\cdots}j_{n}}
\end{eqnarray}
if $D$ is small enough.

\vspace{10mm}
\subsection{Proof for $n=2$}
Let us prove (\ref{eq:fundamental-formula}) for $n=2$.  
From (\ref{eq:quadratic polynomial}) we set for simplicity
\begin{eqnarray}
w&=&F(x,y)=\sum_{i,j=0}^{2}a_{ij}x^{i}y^{j} \nonumber \\
&=&a_{20}x^{2}+a_{10}x+a_{00}+(a_{21}x^{2}+a_{11}x+a_{01})y+
      (a_{22}x^{2}+a_{12}x+a_{02})y^{2}.
\end{eqnarray}
See the figure once more.

\vspace{10mm}
\hspace{-20mm}
 \input{rectangle.fig}

\vspace{10mm} \noindent
Then
\begin{eqnarray}
\label{eq:2-integral}
&&\int_{\alpha_{1}-\Delta_{1}}^{\alpha_{1}+\Delta_{1}}
\int_{\alpha_{2}-\Delta_{2}}^{\alpha_{2}+\Delta_{2}}wdxdy \nonumber \\
&=&
\left\{a_{20}\frac{(\alpha_{1}+\Delta_{1})^{3}-(\alpha_{1}-\Delta_{1})^{3}}{3}
+a_{10}\frac{(\alpha_{1}+\Delta_{1})^{2}-(\alpha_{1}-\Delta_{1})^{2}}{2}
+a_{00}{\cdot}2\Delta_{1}\right\}2\Delta_{2} \nonumber \\
&+&
\left\{a_{21}\frac{(\alpha_{1}+\Delta_{1})^{3}-(\alpha_{1}-\Delta_{1})^{3}}{3}
+a_{11}\frac{(\alpha_{1}+\Delta_{1})^{2}-(\alpha_{1}-\Delta_{1})^{2}}{2}
+a_{01}{\cdot}2\Delta_{1}\right\}\times  \nonumber \\
&&\frac{(\alpha_{2}+\Delta_{2})^{2}-(\alpha_{2}-\Delta_{2})^{2}}{2}
\nonumber \\
&+&
\left\{a_{22}\frac{(\alpha_{1}+\Delta_{1})^{3}-(\alpha_{1}-\Delta_{1})^{3}}{3}
+a_{12}\frac{(\alpha_{1}+\Delta_{1})^{2}-(\alpha_{1}-\Delta_{1})^{2}}{2}
+a_{02}{\cdot}2\Delta_{1}\right\}\times \nonumber \\
&&\frac{(\alpha_{2}+\Delta_{2})^{3}-(\alpha_{2}-\Delta_{2})^{3}}{3}
\nonumber \\
&=&\cdots \nonumber \\
&=&
4\Delta_{1}\Delta_{2}
\left\{
a_{20}\alpha_{1}^{2}+a_{10}\alpha_{1}+a_{00}
+(a_{21}\alpha_{1}^{2}+a_{11}\alpha_{1}+a_{01})\alpha_{2}
+(a_{22}\alpha_{1}^{2}+a_{12}\alpha_{1}+a_{02})\alpha_{2}^{2}
\right\} \nonumber \\
&+&
\frac{4\Delta_{1}\Delta_{2}}{3}
(a_{20}\Delta_{1}^{2}+a_{21}\Delta_{1}^{2}\alpha_{2}+a_{22}\Delta_{1}^{2}\alpha_{2}^{2}
+a_{02}\Delta_{2}^{2}+a_{12}\alpha_{1}\Delta_{2}^{2}+a_{22}\alpha_{1}^{2}\Delta_{2}^{2})
\nonumber \\
&+&
\frac{4\Delta_{1}\Delta_{2}}{9}a_{22}\Delta_{1}^{2}\Delta_{2}^{2}. 
\end{eqnarray}

From (\ref{eq:lattice-points}) we have $9 (=3^{2})$--data
\begin{equation}
w_{k,l}=w(\alpha_{1}+k\Delta_{1},\alpha_{2}+l\Delta_{2})
\end{equation}
for $k,\ l\in \{-1,0,1\}$, so we want to express the integral above 
in terms of these data $\{w_{kl}\}$.

\par \noindent
Since
\begin{equation}
\label{eq:F-1}
a_{20}\alpha_{1}^{2}+a_{10}\alpha_{1}+a_{00}
+(a_{21}\alpha_{1}^{2}+a_{11}\alpha_{1}+a_{01})\alpha_{2}
+(a_{22}\alpha_{1}^{2}+a_{12}\alpha_{1}+a_{02})\alpha_{2}^{2}
=
w_{0,0}
\end{equation}
is clear we must treat the remaining ones
\begin{eqnarray*}
&&a_{20}\Delta_{1}^{2}+a_{21}\Delta_{1}^{2}\alpha_{2}+a_{22}\Delta_{1}^{2}\alpha_{2}^{2}
+a_{02}\Delta_{2}^{2}+a_{12}\alpha_{1}\Delta_{2}^{2}+a_{22}\alpha_{1}^{2}\Delta_{2}^{2}, \\
&&a_{22}\Delta_{1}^{2}\Delta_{2}^{2}.
\end{eqnarray*}

It is not difficult to show
\begin{eqnarray*}
a_{20}\Delta_{1}^{2}+a_{21}\Delta_{1}^{2}\alpha_{2}+a_{22}\Delta_{1}^{2}\alpha_{2}^{2}
&=&
\frac{w_{-1,0}-2w_{0,0}+w_{1,0}}{2},  \\
a_{02}\Delta_{2}^{2}+a_{12}\alpha_{1}\Delta_{2}^{2}+a_{22}\alpha_{1}^{2}\Delta_{2}^{2}
&=&
\frac{w_{0,-1}-2w_{0,0}+w_{0,1}}{2},
\end{eqnarray*}
so we have
\begin{eqnarray}
\label{eq:F-2}
\ &&a_{20}\Delta_{1}^{2}+a_{21}\Delta_{1}^{2}\alpha_{2}+a_{22}\Delta_{1}^{2}\alpha_{2}^{2}
+a_{02}\Delta_{2}^{2}+a_{12}\alpha_{1}\Delta_{2}^{2}+a_{22}\alpha_{1}^{2}\Delta_{2}^{2}
\nonumber \\
&=&
\frac{w_{-1,0}+w_{1,0}+w_{0,-1}+w_{0,1}-4w_{0,0}}{2}.
\end{eqnarray}

Similarly,
\begin{eqnarray*}
\ &&\frac{w_{-1,-1}+w_{-1,1}+w_{1,-1}+w_{1,1}}{4} \\
&=&
a_{20}\alpha_{1}^{2}+a_{10}\alpha_{1}+a_{00}
+(a_{21}\alpha_{1}^{2}+a_{11}\alpha_{1}+a_{01})\alpha_{2}
+(a_{22}\alpha_{1}^{2}+a_{12}\alpha_{1}+a_{02})\alpha_{2}^{2} \\
&+&
a_{20}\Delta_{1}^{2}+a_{21}\Delta_{1}^{2}\alpha_{2}+a_{22}\Delta_{1}^{2}\alpha_{2}^{2}
+a_{02}\Delta_{2}^{2}+a_{12}\alpha_{1}\Delta_{2}^{2}+a_{22}\alpha_{1}^{2}\Delta_{2}^{2}
\\
&+& a_{22}\Delta_{1}^{2}\Delta_{2}^{2} \\
&=&
w_{0,0}+\frac{w_{-1,0}+w_{1,0}+w_{0,-1}+w_{0,1}-4w_{0,0}}{2}+
a_{22}\Delta_{1}^{2}\Delta_{2}^{2}
\end{eqnarray*}
so we have
\begin{equation}
\label{eq:F-3}
a_{22}\Delta_{1}^{2}\Delta_{2}^{2}
=
\frac{w_{-1,-1}+w_{-1,1}+w_{1,-1}+w_{1,1}
-2(w_{-1,0}+w_{1,0}+w_{0,-1}+w_{0,1})+4w_{0,0}}{4}.
\end{equation}

Therefore, substituting (\ref{eq:F-1}), (\ref{eq:F-2}) and 
 (\ref{eq:F-3}) into (\ref{eq:2-integral}) we obtain
\begin{eqnarray}
\label{eq:formula-2}
\ &&\int_{\alpha_{1}-\Delta_{1}}^{\alpha_{1}+\Delta_{1}}
\int_{\alpha_{2}-\Delta_{2}}^{\alpha_{2}+\Delta_{2}}wdxdy
\nonumber \\
&=&
4\Delta_{1}\Delta_{2}w_{0,0}
+
\frac{2\Delta_{1}\Delta_{2}}{3}(w_{-1,0}+w_{1,0}+w_{0,-1}+w_{0,1}-4w_{0,0}) 
\nonumber \\
&+&
\frac{\Delta_{1}\Delta_{2}}{9}
(w_{-1,-1}+w_{-1,1}+w_{1,-1}+w_{1,1}-2(w_{-1,0}+w_{1,0}+w_{0,-1}+w_{0,1})+4w_{0,0})
\nonumber \\
&=&
\frac{\Delta_{1}\Delta_{2}}{9}
\left\{
w_{-1,-1}+w_{-1,1}+w_{1,-1}+w_{1,1}+4(w_{-1,0}+w_{1,0}+w_{0,-1}+w_{0,1})+16w_{0,0}
\right\}. \nonumber \\
&&
\end{eqnarray}

This formula is interesting enough.

\subsection{Proof for $n=3$}
Let us prove (\ref{eq:fundamental-formula}) for $n=3$. 
From (\ref{eq:quadratic polynomial}) we set
\begin{eqnarray}
w&=&F(x,y,z)=\sum_{i,j,k=0}^{2}a_{ijk}x^{i}y^{j}z^{k} \nonumber \\
&=&
a_{200}x^{2}+a_{100}x+a_{000}+(a_{210}x^{2}+a_{110}x+a_{010})y+
(a_{220}x^{2}+a_{120}x+a_{020})y^{2}  \nonumber \\
&+&
\left\{
a_{201}x^{2}+a_{101}x+a_{001}+(a_{211}x^{2}+a_{111}x+a_{011})y+
(a_{221}x^{2}+a_{121}x+a_{021})y^{2}
\right\}z  \nonumber \\
&+&
\left\{
a_{202}x^{2}+a_{102}x+a_{002}+(a_{212}x^{2}+a_{112}x+a_{012})y+
(a_{222}x^{2}+a_{122}x+a_{022})y^{2}
\right\}z^{2}.  \nonumber \\
&&
\end{eqnarray}
Then
\begin{eqnarray*}
\label{eq:3-integral}
&&\int_{\alpha_{1}-\Delta_{1}}^{\alpha_{1}+\Delta_{1}}
\int_{\alpha_{2}-\Delta_{2}}^{\alpha_{2}+\Delta_{2}}
\int_{\alpha_{3}-\Delta_{3}}^{\alpha_{3}+\Delta_{3}}
wdxdydz  \nonumber \\
&=&
\left\{
\int_{\alpha_{1}-\Delta_{1}}^{\alpha_{1}+\Delta_{1}}
\int_{\alpha_{2}-\Delta_{2}}^{\alpha_{2}+\Delta_{2}}
\sum_{i,j=0}^{2}a_{ij0}x^{i}y^{j}
\right\}
2\Delta_{3}  \\
&+&
\left\{
\int_{\alpha_{1}-\Delta_{1}}^{\alpha_{1}+\Delta_{1}}
\int_{\alpha_{2}-\Delta_{2}}^{\alpha_{2}+\Delta_{2}}
\sum_{i,j=0}^{2}a_{ij1}x^{i}y^{j}
\right\}
\frac{(\alpha_{3}+\Delta_{3})^{2}-(\alpha_{3}-\Delta_{3})^{2}}{2}  \\
&+&
\left\{
\int_{\alpha_{1}-\Delta_{1}}^{\alpha_{1}+\Delta_{1}}
\int_{\alpha_{2}-\Delta_{2}}^{\alpha_{2}+\Delta_{2}}
\sum_{i,j=0}^{2}a_{ij2}x^{i}y^{j}
\right\}
\frac{(\alpha_{3}+\Delta_{3})^{3}-(\alpha_{3}-\Delta_{3})^{3}}{3}.
\end{eqnarray*}
By (\ref{eq:2-integral}) some calculation gives

\begin{eqnarray}
&&\int_{\alpha_{1}-\Delta_{1}}^{\alpha_{1}+\Delta_{1}}
\int_{\alpha_{2}-\Delta_{2}}^{\alpha_{2}+\Delta_{2}}
\int_{\alpha_{3}-\Delta_{3}}^{\alpha_{3}+\Delta_{3}}
wdxdydz  \nonumber \\
&=&
[
4\Delta_{1}\Delta_{2}
\{
a_{200}\alpha_{1}^{2}+a_{100}\alpha_{1}+a_{000}
+(a_{210}\alpha_{1}^{2}+a_{110}\alpha_{1}+a_{010})\alpha_{2}
+(a_{220}\alpha_{1}^{2}+a_{120}\alpha_{1}+a_{020})\alpha_{2}^{2}
\}  \nonumber \\
&+&
\frac{4\Delta_{1}\Delta_{2}}{3}
(a_{200}\Delta_{1}^{2}+a_{210}\Delta_{1}^{2}\alpha_{2}+a_{220}\Delta_{1}^{2}\alpha_{2}^{2}
+a_{020}\Delta_{2}^{2}+a_{120}\alpha_{1}\Delta_{2}^{2}+a_{220}\alpha_{1}^{2}\Delta_{2}^{2})
\nonumber \\
&+&
\frac{4\Delta_{1}\Delta_{2}}{9}a_{220}\Delta_{1}^{2}\Delta_{2}^{2}
]2\Delta_{3}  \nonumber \\
&+&
[
4\Delta_{1}\Delta_{2}
\{
a_{201}\alpha_{1}^{2}+a_{101}\alpha_{1}+a_{001}
+(a_{211}\alpha_{1}^{2}+a_{111}\alpha_{1}+a_{011})\alpha_{2}
+(a_{221}\alpha_{1}^{2}+a_{121}\alpha_{1}+a_{021})\alpha_{2}^{2}
\}  \nonumber \\
&+&
\frac{4\Delta_{1}\Delta_{2}}{3}
(a_{201}\Delta_{1}^{2}+a_{211}\Delta_{1}^{2}\alpha_{2}+a_{221}\Delta_{1}^{2}\alpha_{2}^{2}
+a_{021}\Delta_{2}^{2}+a_{121}\alpha_{1}\Delta_{2}^{2}+a_{221}\alpha_{1}^{2}\Delta_{2}^{2})
\nonumber \\
&+&
\frac{4\Delta_{1}\Delta_{2}}{9}a_{221}\Delta_{1}^{2}\Delta_{2}^{2}
]2\alpha_{3}\Delta_{3}  \nonumber \\
&+&
[
4\Delta_{1}\Delta_{2}
\{
a_{202}\alpha_{1}^{2}+a_{102}\alpha_{1}+a_{002}
+(a_{212}\alpha_{1}^{2}+a_{112}\alpha_{1}+a_{012})\alpha_{2}
+(a_{222}\alpha_{1}^{2}+a_{122}\alpha_{1}+a_{022})\alpha_{2}^{2}
\}  \nonumber \\
&+&
\frac{4\Delta_{1}\Delta_{2}}{3}
(a_{202}\Delta_{1}^{2}+a_{212}\Delta_{1}^{2}\alpha_{2}+a_{222}\Delta_{1}^{2}\alpha_{2}^{2}
+a_{022}\Delta_{2}^{2}+a_{122}\alpha_{1}\Delta_{2}^{2}+a_{222}\alpha_{1}^{2}\Delta_{2}^{2})
\nonumber \\
&+&
\frac{4\Delta_{1}\Delta_{2}}{9}a_{222}\Delta_{1}^{2}\Delta_{2}^{2}
](2\alpha_{3}^{2}\Delta_{3}+\frac{2}{3}\Delta_{3}^{3})  \nonumber \\
&=&
8\Delta_{1}\Delta_{2}\Delta_{3}\times  \nonumber \\
&&  
[\ 
\{
a_{200}\alpha_{1}^{2}+a_{100}\alpha_{1}+a_{000}
+(a_{210}\alpha_{1}^{2}+a_{110}\alpha_{1}+a_{010})\alpha_{2}
+(a_{220}\alpha_{1}^{2}+a_{120}\alpha_{1}+a_{020})\alpha_{2}^{2}
\}  \nonumber \\
&&+
\{
a_{201}\alpha_{1}^{2}+a_{101}\alpha_{1}+a_{001}
+(a_{211}\alpha_{1}^{2}+a_{111}\alpha_{1}+a_{011})\alpha_{2}
+(a_{221}\alpha_{1}^{2}+a_{121}\alpha_{1}+a_{021})\alpha_{2}^{2}
\}\alpha_{3}  \nonumber \\
&&+
\{
a_{202}\alpha_{1}^{2}+a_{102}\alpha_{1}+a_{002}
+(a_{212}\alpha_{1}^{2}+a_{112}\alpha_{1}+a_{012})\alpha_{2}
+(a_{222}\alpha_{1}^{2}+a_{122}\alpha_{1}+a_{022})\alpha_{2}^{2}
\}\alpha_{3}^{2}
] \nonumber \\
&+&
\frac{8\Delta_{1}\Delta_{2}\Delta_{3}}{3}\times  \nonumber \\
&&
[\ 
\{
a_{200}\Delta_{1}^{2}+a_{210}\Delta_{1}^{2}\alpha_{2}+a_{220}\Delta_{1}^{2}\alpha_{2}^{2}
+a_{020}\Delta_{2}^{2}+a_{120}\alpha_{1}\Delta_{2}^{2}+a_{220}\alpha_{1}^{2}\Delta_{2}^{2}
\}  \nonumber \\
&&+
\{
a_{201}\Delta_{1}^{2}+a_{211}\Delta_{1}^{2}\alpha_{2}+a_{221}\Delta_{1}^{2}\alpha_{2}^{2}
+a_{021}\Delta_{2}^{2}+a_{121}\alpha_{1}\Delta_{2}^{2}+a_{221}\alpha_{1}^{2}\Delta_{2}^{2}
\}\alpha_{3}  \nonumber \\
&&+
\{
a_{202}\Delta_{1}^{2}+a_{212}\Delta_{1}^{2}\alpha_{2}+a_{222}\Delta_{1}^{2}\alpha_{2}^{2}
+a_{022}\Delta_{2}^{2}+a_{122}\alpha_{1}\Delta_{2}^{2}+a_{222}\alpha_{1}^{2}\Delta_{2}^{2}
\}\alpha_{3}^{2}
]  \nonumber \\
&+&
\frac{8\Delta_{1}\Delta_{2}\Delta_{3}}{9}
(a_{220}\Delta_{1}^{2}\Delta_{2}^{2}+
a_{221}\Delta_{1}^{2}\Delta_{2}^{2}\alpha_{3}+
a_{222}\Delta_{1}^{2}\Delta_{2}^{2}\alpha_{3}^{2}) \nonumber \\
&+&
\frac{8\Delta_{1}\Delta_{2}\Delta_{3}}{3}\times \nonumber \\
&&
\{
a_{202}\alpha_{1}^{2}+a_{102}\alpha_{1}+a_{002}
+(a_{212}\alpha_{1}^{2}+a_{112}\alpha_{1}+a_{012})\alpha_{2}
+(a_{222}\alpha_{1}^{2}+a_{122}\alpha_{1}+a_{022})\alpha_{2}^{2}
\}\Delta_{3}^{2} \nonumber \\
&+&
\frac{8\Delta_{1}\Delta_{2}\Delta_{3}}{9}
(a_{202}\Delta_{1}^{2}+a_{212}\Delta_{1}^{2}\alpha_{2}+a_{222}\Delta_{1}^{2}\alpha_{2}^{2}
+a_{022}\Delta_{2}^{2}+a_{122}\alpha_{1}\Delta_{2}^{2}+a_{222}\alpha_{1}^{2}\Delta_{2}^{2})
\Delta_{3}^{2} \nonumber \\
&+&
\frac{8\Delta_{1}\Delta_{2}\Delta_{3}}{27}
a_{222}\Delta_{1}^{2}\Delta_{2}^{2}\Delta_{3}^{2}.  \nonumber
\end{eqnarray}

By rearranging terms we have
\begin{eqnarray}
\label{eq:3-integral}
&&\int_{\alpha_{1}-\Delta_{1}}^{\alpha_{1}+\Delta_{1}}
\int_{\alpha_{2}-\Delta_{2}}^{\alpha_{2}+\Delta_{2}}
\int_{\alpha_{3}-\Delta_{3}}^{\alpha_{3}+\Delta_{3}}
wdxdydz  \nonumber \\
&=&
8\Delta_{1}\Delta_{2}\Delta_{3}\times  \nonumber \\
&&  
[\ 
\{
a_{200}\alpha_{1}^{2}+a_{100}\alpha_{1}+a_{000}
+(a_{210}\alpha_{1}^{2}+a_{110}\alpha_{1}+a_{010})\alpha_{2}
+(a_{220}\alpha_{1}^{2}+a_{120}\alpha_{1}+a_{020})\alpha_{2}^{2}
\}  \nonumber \\
&&+
\{
a_{201}\alpha_{1}^{2}+a_{101}\alpha_{1}+a_{001}
+(a_{211}\alpha_{1}^{2}+a_{111}\alpha_{1}+a_{011})\alpha_{2}
+(a_{221}\alpha_{1}^{2}+a_{121}\alpha_{1}+a_{021})\alpha_{2}^{2}
\}\alpha_{3}  \nonumber \\
&&+
\{
a_{202}\alpha_{1}^{2}+a_{102}\alpha_{1}+a_{002}
+(a_{212}\alpha_{1}^{2}+a_{112}\alpha_{1}+a_{012})\alpha_{2}
+(a_{222}\alpha_{1}^{2}+a_{122}\alpha_{1}+a_{022})\alpha_{2}^{2}
\}\alpha_{3}^{2}
] \nonumber \\
&+&
\frac{8\Delta_{1}\Delta_{2}\Delta_{3}}{3}\times  \nonumber \\
&&
[\
\{
a_{200}+a_{210}\alpha_{2}+a_{220}\alpha_{2}^{2}
+(a_{201}+a_{211}\alpha_{2}+a_{221}\alpha_{2}^{2})\alpha_{3}
+(a_{202}+a_{212}\alpha_{2}+a_{222}\alpha_{2}^{2})\alpha_{3}^{2}
\}\Delta_{1}^{2}  \nonumber \\
&&+ 
\{
a_{020}+a_{120}\alpha_{1}+a_{220}\alpha_{1}^{2}
+(a_{021}+a_{121}\alpha_{1}+a_{221}\alpha_{1}^{2})\alpha_{3}
+(a_{022}+a_{122}\alpha_{1}+a_{222}\alpha_{1}^{2})\alpha_{3}^{2}
\}\Delta_{2}^{2}  \nonumber \\
&&+
\{
a_{002}+a_{102}\alpha_{1}+a_{202}\alpha_{1}^{2}
+(a_{012}+a_{112}\alpha_{1}+a_{212}\alpha_{1}^{2})\alpha_{2}
+(a_{022}+a_{122}\alpha_{1}+a_{222}\alpha_{1}^{2})\alpha_{2}^{2}
\}\Delta_{3}^{2}
]  \nonumber \\
&+&
\frac{8\Delta_{1}\Delta_{2}\Delta_{3}}{9}\times  \nonumber \\
&&
\{
a_{220}\Delta_{1}^{2}\Delta_{2}^{2}+
a_{221}\Delta_{1}^{2}\Delta_{2}^{2}\alpha_{3}+
a_{222}\Delta_{1}^{2}\Delta_{2}^{2}\alpha_{3}^{2}+
a_{202}\Delta_{1}^{2}\Delta_{3}^{2}+
a_{212}\Delta_{1}^{2}\alpha_{2}\Delta_{3}^{2}+
a_{222}\Delta_{1}^{2}\alpha_{2}^{2}\Delta_{3}^{2}
\nonumber \\
&&+
a_{022}\Delta_{2}^{2}\Delta_{3}^{2}+
a_{122}\alpha_{1}\Delta_{2}^{2}\Delta_{3}^{2}+
a_{222}\alpha_{1}^{2}\Delta_{2}^{2}\Delta_{3}^{2}
\} \nonumber \\
&+&
\frac{8\Delta_{1}\Delta_{2}\Delta_{3}}{27}
a_{222}\Delta_{1}^{2}\Delta_{2}^{2}\Delta_{3}^{2}.
\end{eqnarray}

From (\ref{eq:lattice-points}) we have $27 (=3^{3})$--data
\begin{equation}
w_{l,m,n}=w(\alpha_{1}+l\Delta_{1},\alpha_{2}+m\Delta_{2},\alpha_{3}+n\Delta_{3})
\end{equation}
for $l, m, n \in \{-1,0,1\}$, so we must express the integral above 
in terms of these data $\{w_{l,m,n}\}$. 
Since
\begin{eqnarray}
\label{eq:K-1}
&&\ \
\{
a_{200}\alpha_{1}^{2}+a_{100}\alpha_{1}+a_{000}
+(a_{210}\alpha_{1}^{2}+a_{110}\alpha_{1}+a_{010})\alpha_{2}
+(a_{220}\alpha_{1}^{2}+a_{120}\alpha_{1}+a_{020})\alpha_{2}^{2}
\}  \nonumber \\
&&+
\{
a_{201}\alpha_{1}^{2}+a_{101}\alpha_{1}+a_{001}
+(a_{211}\alpha_{1}^{2}+a_{111}\alpha_{1}+a_{011})\alpha_{2}
+(a_{221}\alpha_{1}^{2}+a_{121}\alpha_{1}+a_{021})\alpha_{2}^{2}
\}\alpha_{3}  \nonumber \\
&&+
\{
a_{202}\alpha_{1}^{2}+a_{102}\alpha_{1}+a_{002}
+(a_{212}\alpha_{1}^{2}+a_{112}\alpha_{1}+a_{012})\alpha_{2}
+(a_{222}\alpha_{1}^{2}+a_{122}\alpha_{1}+a_{022})\alpha_{2}^{2}
\}\alpha_{3}^{2}  \nonumber \\
&&= w_{0,0,0}
\end{eqnarray}
we treat the remaining ones. Here we list formulas necessary to prove 
the main formula. 

\begin{eqnarray}
\label{eq:K-2}
&&\ \
\{
a_{200}+a_{210}\alpha_{2}+a_{220}\alpha_{2}^{2}
+(a_{201}+a_{211}\alpha_{2}+a_{221}\alpha_{2}^{2})\alpha_{3}
+(a_{202}+a_{212}\alpha_{2}+a_{222}\alpha_{2}^{2})\alpha_{3}^{2}
\}\Delta_{1}^{2}  \nonumber \\
&&=
\frac{1}{2}(w_{-1,0,0}+w_{1,0,0})-w_{0,0,0}, \nonumber \\
&&\ \ 
\{
a_{020}+a_{120}\alpha_{1}+a_{220}\alpha_{1}^{2}
+(a_{021}+a_{121}\alpha_{1}+a_{221}\alpha_{1}^{2})\alpha_{3}
+(a_{022}+a_{122}\alpha_{1}+a_{222}\alpha_{1}^{2})\alpha_{3}^{2}
\}\Delta_{2}^{2}  \nonumber \\
&&=
\frac{1}{2}(w_{0,-1,0}+w_{0,1,0})-w_{0,0,0}, \nonumber \\
&&\ \ 
\{
a_{002}+a_{102}\alpha_{1}+a_{202}\alpha_{1}^{2}
+(a_{012}+a_{112}\alpha_{1}+a_{212}\alpha_{1}^{2})\alpha_{2}
+(a_{022}+a_{122}\alpha_{1}+a_{222}\alpha_{1}^{2})\alpha_{2}^{2}
\}\Delta_{3}^{2}  \nonumber \\
&&=
\frac{1}{2}(w_{0,0,-1}+w_{0,0,1})-w_{0,0,0}
\end{eqnarray}
and
\begin{eqnarray}
\label{eq:K-3}
&&\ \ 
a_{220}\Delta_{1}^{2}\Delta_{2}^{2}+
a_{221}\Delta_{1}^{2}\Delta_{2}^{2}\alpha_{3}+
a_{222}\Delta_{1}^{2}\Delta_{2}^{2}\alpha_{3}^{2}+
a_{202}\Delta_{1}^{2}\Delta_{3}^{2}+
a_{212}\Delta_{1}^{2}\alpha_{2}\Delta_{3}^{2}+
a_{222}\Delta_{1}^{2}\alpha_{2}^{2}\Delta_{3}^{2}
\nonumber \\
&&+
a_{022}\Delta_{2}^{2}\Delta_{3}^{2}+
a_{122}\alpha_{1}\Delta_{2}^{2}\Delta_{3}^{2}+
a_{222}\alpha_{1}^{2}\Delta_{2}^{2}\Delta_{3}^{2}
\nonumber \\
&&=
\frac{1}{4}(w_{-1,-1,0}+w_{-1,1,0}+w_{1,-1,0}+w_{1,1,0}
+w_{-1,0,-1}+w_{-1,0,1}+w_{1,0,-1}+w_{1,0,1}  \nonumber \\
&&\qquad
+w_{0,-1,-1}+w_{0,-1,1}+w_{0,1,-1}+w_{0,1,1})  \nonumber \\
&&\ \ 
-(w_{-1,0,0}+w_{1,0,0}+w_{0,-1,0}+w_{0,1,0}+w_{0,0,-1}+w_{0,0,1})+3w_{0,0,0} 
\end{eqnarray}
and
\begin{eqnarray}
\label{eq:K-4}
&&
a_{222}\Delta_{1}^{2}\Delta_{2}^{2}\Delta_{3}^{2}  \nonumber \\
&&=
\frac{1}{8}(w_{-1,-1,-1}+w_{1,-1,-1}+w_{1,1,-1}+w_{-1,1,-1}
+w_{-1,-1,1}+w_{1,-1,1}+w_{1,1,1}+w_{-1,1,1})  \nonumber \\
&&\ 
-\frac{1}{4}(w_{-1,-1,0}+w_{-1,1,0}+w_{1,-1,0}+w_{1,1,0}
+w_{-1,0,-1}+w_{-1,0,1}+w_{1,0,-1}+w_{1,0,1} \nonumber \\
&&\qquad 
+w_{0,-1,-1}+w_{0,-1,1}+w_{0,1,-1}+w_{0,1,1})  \nonumber \\
&&\ 
+\frac{1}{2}(w_{-1,0,0}+w_{1,0,0}+w_{0,-1,0}+w_{0,1,0}
+w_{0,0,-1}+w_{0,0,1})-w_{0,0,0}.
\end{eqnarray}
The proofs are long but straightforward, so they are left to 
readers.

Therefore, substituting (\ref{eq:K-1}) $\sim$ (\ref{eq:K-4}) into 
(\ref{eq:3-integral}) we obtain

\begin{eqnarray}
\label{eq:formula-3}
&&\int_{\alpha_{1}-\Delta_{1}}^{\alpha_{1}+\Delta_{1}}
\int_{\alpha_{2}-\Delta_{2}}^{\alpha_{2}+\Delta_{2}}
\int_{\alpha_{3}-\Delta_{3}}^{\alpha_{3}+\Delta_{3}}
wdxdydz  \nonumber \\
&=&
8\Delta_{1}\Delta_{2}\Delta_{3}\ w_{0,0,0}  \nonumber \\
&+&
\frac{8\Delta_{1}\Delta_{2}\Delta_{3}}{3}
\left\{
\frac{1}{2}(w_{-1,0,0}+w_{1,0,0}+w_{0,-1,0}+w_{0,1,0}+w_{0,0,-1}+w_{0,0,1})
-3w_{0,0,0}
\right\}  \nonumber \\
&+&
\frac{8\Delta_{1}\Delta_{2}\Delta_{3}}{9}
\left\{
\frac{1}{4}(w_{-1,-1,0}+w_{-1,1,0}+w_{1,-1,0}+w_{1,1,0}
+w_{-1,0,-1}+w_{-1,0,1}+w_{1,0,-1}+w_{1,0,1} 
\right.  \nonumber \\
&&\ 
+w_{0,-1,-1}+w_{0,-1,1}+w_{0,1,-1}+w_{0,1,1}) \nonumber \\
&&\left. \ 
-(w_{-1,0,0}+w_{1,0,0}+w_{0,-1,0}+w_{0,1,0}+w_{0,0,-1}+w_{0,0,1})+3w_{0,0,0} 
\right\} \nonumber \\
&+&
\frac{8\Delta_{1}\Delta_{2}\Delta_{3}}{27}
\left\{
\frac{1}{8}(w_{-1,-1,-1}+w_{1,-1,-1}+w_{1,1,-1}+w_{-1,1,-1}
+w_{-1,-1,1}+w_{1,-1,1}+w_{1,1,1}+w_{-1,1,1})  
\right. \nonumber \\
&&\ 
-\frac{1}{4}(w_{-1,-1,0}+w_{-1,1,0}+w_{1,-1,0}+w_{1,1,0}
+w_{-1,0,-1}+w_{-1,0,1}+w_{1,0,-1}+w_{1,0,1} \nonumber \\
&&\quad\ \ 
+w_{0,-1,-1}+w_{0,-1,1}+w_{0,1,-1}+w_{0,1,1})  \nonumber \\
&&\left. 
+\frac{1}{2}(w_{-1,0,0}+w_{1,0,0}+w_{0,-1,0}+w_{0,1,0}
+w_{0,0,-1}+w_{0,0,1})-w_{0,0,0}
\right\}  \nonumber \\
&=&
\frac{\Delta_{1}\Delta_{2}\Delta_{3}}{27}\times  \nonumber \\
&&
\left\{
w_{-1,-1,-1}+w_{1,-1,-1}+w_{1,1,-1}+w_{-1,1,-1}
+w_{-1,-1,1}+w_{1,-1,1}+w_{1,1,1}+w_{-1,1,1}  
\right. \nonumber \\
&&+
4(w_{-1,-1,0}+w_{-1,1,0}+w_{1,-1,0}+w_{1,1,0}
+w_{-1,0,-1}+w_{-1,0,1}+w_{1,0,-1}+w_{1,0,1} \nonumber \\
&&\quad\
+w_{0,-1,-1}+w_{0,-1,1}+w_{0,1,-1}+w_{0,1,1})  \nonumber \\
&&+
16(w_{-1,0,0}+w_{1,0,0}+w_{0,-1,0}+w_{0,1,0}
+w_{0,0,-1}+w_{0,0,1})  \nonumber \\
&&
\left. 
+64w_{0,0,0}
\right\}.
\end{eqnarray}

The proof is a bit complicated.

\subsection{General Case}
Let us consider the general case.  Unfortunately, we have 
no method to calculate the general case at the present time. 
However, from the results  (\ref{eq:formula-2}) and (\ref{eq:formula-3}) 
we can conjecture the general formula as follows. 

The formulas (\ref{eq:formula-2}) and (\ref{eq:formula-3}) can be 
rewritten as
\begin{eqnarray*}
\ &&\int_{\alpha_{1}-\Delta_{1}}^{\alpha_{1}+\Delta_{1}}
\int_{\alpha_{2}-\Delta_{2}}^{\alpha_{2}+\Delta_{2}}wdxdy  \\
&=&
\frac{\Delta_{1}\Delta_{2}}{3^{2}}
\left\{
w_{-1,-1}+w_{-1,1}+w_{1,-1}+w_{1,1}+4(w_{-1,0}+w_{1,0}+w_{0,-1}+w_{0,1})+4^{2}w_{0,0}
\right\}
\end{eqnarray*}
and
\begin{eqnarray*}
&&\int_{\alpha_{1}-\Delta_{1}}^{\alpha_{1}+\Delta_{1}}
\int_{\alpha_{2}-\Delta_{2}}^{\alpha_{2}+\Delta_{2}}
\int_{\alpha_{3}-\Delta_{3}}^{\alpha_{3}+\Delta_{3}}
wdxdydz  \\
&=&
\frac{\Delta_{1}\Delta_{2}\Delta_{3}}{3^{3}}\times  \nonumber \\
&&
\left\{
w_{-1,-1,-1}+w_{1,-1,-1}+w_{1,1,-1}+w_{-1,1,-1}
+w_{-1,-1,1}+w_{1,-1,1}+w_{1,1,1}+w_{-1,1,1}  
\right.  \\
&&+
4(w_{-1,-1,0}+w_{-1,1,0}+w_{1,-1,0}+w_{1,1,0}
+w_{-1,0,-1}+w_{-1,0,1}+w_{1,0,-1}+w_{1,0,1}  \\
&&\quad\
+w_{0,-1,-1}+w_{0,-1,1}+w_{0,1,-1}+w_{0,1,1})  \\
&&+
4^{2}(w_{-1,0,0}+w_{1,0,0}+w_{0,-1,0}+w_{0,1,0}
+w_{0,0,-1}+w_{0,0,1})  \\
&&\left. 
+4^{3}w_{0,0,0}
\right\}.
\end{eqnarray*}
From the forms it is easy to conjecture the general form. 

\vspace{3mm}
If we set
\begin{equation}
w=F(x_{1}, x_{2}, \cdots , x_{n})
=\sum_{i_{1},i_{2},\cdots,i_{n}=0}^{2}a_{i_{1}i_{2}\cdots i_{n}}
  x_{1}^{i_{1}}x_{2}^{i_{2}}\cdots x_{n}^{i_{n}}
\end{equation}
then
\begin{eqnarray}
\label{eq:formula-n}
&&\int_{\alpha_{1}-\Delta_{1}}^{\alpha_{1}+\Delta_{1}}
\int_{\alpha_{2}-\Delta_{2}}^{\alpha_{2}+\Delta_{2}}
\cdots 
\int_{\alpha_{n}-\Delta_{n}}^{\alpha_{n}+\Delta_{n}}
wdx_{1}dx_{2} \cdots dx_{n}  \nonumber \\
&& \nonumber \\
&&=\frac{\Delta_{1}\Delta_{2}\cdots \Delta_{n}}{3^{n}}
      \sum_{ j_{1},j_{2},\cdots ,j_{n} \in \{-1,0,1\} }
      4^{\sharp\{j_{1},j_{2},\cdots ,j_{n}\}_{0}}w_{j_{1}j_{2}{\cdots}j_{n}}
\end{eqnarray}
where $\sharp\{j_{1},j_{2},\cdots ,j_{n}\}_{0}$ is the number of $0$ in 
$\{j_{1},j_{2},\cdots ,j_{n}\}$ and 
\begin{eqnarray*}
w_{j_{1}j_{2}{\cdots}j_{n}}
&=&
F(\alpha_{1}+j_{1}\Delta_{1},\alpha_{2}+j_{2}\Delta_{2},\cdots, \alpha_{n}+j_{n}\Delta_{n})
\\
&=&
f(\alpha_{1}+j_{1}\Delta_{1},\alpha_{2}+j_{2}\Delta_{2},\cdots, \alpha_{n}+j_{n}\Delta_{n}).
\end{eqnarray*}

The formula is beautiful enough.

\section{Concluding Remarks}
In the paper we gave a multidimensional analogue of 
the Simpson's formula of integral.  As far as we know the formula 
has not been given, which is a bit mysterious. 
We believe that it will become useful in all fields related to integrals.

Next let us consider more general problems. Let $(M, g)$ be a 
curved space--time and $f$ be a function on $M$. 
If we consider the integral
\[
\int_{M}f(x)dv_{g}(x)
\]
where $dv_{g}$ is some measure on $M$, it is almost impossible 
to calculate.  For that we want to discretize the integral as 
follows.
\[
\sum_{\{D\}}\int_{D}f(x)dv_{g}(x)
\ \longrightarrow\ 
\sum_{\{D\}}\{\mbox{some formula like}\ (\ref{eq:formula-n})\}
\]
Therefore we must choose $D$ to be ``computable", which is just 
key point.

Details of calculation will be reported in \cite{KF}.


\end{document}